\documentclass[12pt]{spieman}  
\usepackage{amsmath,amsfonts,amssymb}
\usepackage{multirow}
\usepackage{graphicx}
\usepackage{setspace}
\usepackage{tocloft}
\usepackage{lineno}
\newcommand{\tcr}{}

\graphicspath{{./}{Figures/}{AP_figs/}}

\title{The Simultaneous Operation of a Controllable Segmented Primary Mirror and Single Conjugate Adaptive Optics System part 1 - Design Concept and Sensitivity Analysis}

\author[a, *]{Benjamin Calvin} 
\author[a]{Michael P. Fitzgerald} 
\affil[a]{University of California - Los Angeles, Physics and Astronomy, 430 Portola Plaza, Los Angeles, CA, USA, 90095}

\cftpagenumbersoff{figure}
\cftpagenumbersoff{table} 
\begin{document} 
\maketitle

\begin{abstract}
The maintenance of primary mirror segment co-phasing is a critical aspect to the operation of segmented telescopes. However, speckle-based measurements of the phasing of the Keck primary have estimated semi-static surface aberrations of approximately 65\,nm rms, 
which were not sensed by the current phasing control system. We propose directly sensing and controlling the primary via the adaptive optics (AO) system, as a Controllable Segmented Primary (CSP), to actively correct its phasing. We develop a methodology for separating the independent controllable signal of the CSP from the rest of the AO system and show estimations of the achievable measurement precision from models of the Keck-II AO system. 

\end{abstract}


{\noindent \footnotesize\textbf{*}Benjamin Calvin,  \linkable{bcalvin@astro.ucla.edu}, \linkable{ben.a.calv@gmail.com} }


\section{Introduction} \label{sec: intro}

Segmented telescopes are critical to both current and future optical and near-IR astronomy. 
Not only are many 
of the largest single-aperture telescopes in operation today segmented, but the planned extremely large class of telescopes all utilize segmented primary designs as well. 
Many of the scientific goals of these large telescopes require a stable response function on the science detector. Astrometric measurements of crowded fields, such as the Galactic Center \cite{2019ApJ...873....9J}, 
consistent coupling into optical fibers for fiber-fed spectroscopy \cite{2017ApJ...838...92M, 2021JATIS...7c5006D}, high-contrast imaging techniques \cite{2016PASP..128j2001B}, and many other applications all require stable 
image quality to be successful. 
In ground-based telescopes with large apertures, adaptive optics (AO) systems are commonly used to remove the time-varying phase distortions caused by the atmosphere. 
However, for a segmented telescope with significant misalignments present in the phasing of the primary mirror, 
the adaptive optics system is unable to perfectly correct the discontinuous phase aberrations that are induced by the physical movement of the primary mirror segments. 
As such, systems that actively maintain the alignment of segmented mirrors are 
crucial for the success of the telescopes' science missions.

Every large ground-based segmented telescope and all of the planned extremely large telescopes require active phasing. The telescopes with closely packed hexagonal segment architectures all use or plan to use capacitance-based edge sensor systems to actively control the alignment of the primary mirror \cite{1990SPIE.1236.1038C, 2000SPIE.4003..355S, 2007Msngr.127...11G, 2008SPIE.7018E..0SB, 2008SPIE.7019E..0IL, 2016SPIE.9906E..06H}. %
%
The alignment precision of the segmented primary 
is fundamentally limited by the electrical noise in the sensors and actuators. 
For the system implemented in the W. M. Keck telescopes, for example, the edge sensors sustain an uncertainty of 5\,nm 
in the measurement of relative displacement between any two contiguous segments. Reference \citenum{1994SPIE.2199..105C} states that this sensor noise propagates across the primary to create an expected uncertainty in the alignment of 27\,nm rms across the primary, 
which primarily manifest as a combination of low-order global Zernike modes \cite{1998SPIE.3352..307T, 2004ApOpt..43.1223C}. 
The observed alignment on the Keck primary mirrors, however, does not meet this target. 
References \citenum{2018SPIE10700E..1DR, 2022SPIE12182E..09R} used speckle-based measurements of the Keck-II primary phasing with NIRC2 to estimate that the surface of the primary mirror is typically misaligned with approximately 65\,nm rms of periodic segment piston errors, which occur at much higher amplitudes and spatial frequencies than the expected residual surface error from from edge sensing. Reference \citenum{2024ApJ...967..171S} provides further evidence that the surface of the Keck-II primary is subject to segment piston errors on magnitudes and spatial distributions that go beyond the expectations of the residual surface errors from edge sensing. References \citenum{2022SPIE12182E..09R, 2023ApOpt..62.5982C, 2024ApJ...967..171S} provide evidence that these surface aberrations evolve as a function of the telescope elevation angle, which would generate semi-static speckles that compromise the stability of the PSF over the course of extended observations. 
Therefore, to maintain high wavefront quality and PSF stability, there must be some auxiliary sensor to help monitor the alignment of the primary. In real systems, static surface aberrations on the primary mirror segments can create ambiguity in when the primary mirror is "properly phased," so, for the purposes of this project, we consider the primary mirror to be aligned when the deviation of any of the actively controlled degrees of freedom of any segment would decrease the Strehl ratio at the focal plane of interest.

Several studies have proposed using a dedicated downstream wavefront sensor for monitoring the phase of the primary \cite{1988ESOC...30..421C, 2005JOSAA..22.1093Y, 2006SPIE.6267E..32M, 2008SPIE.7012E..0ZG}. 
For its capability 
to directly measure phase discontinuities, characteristic of segment misalignments on the primary, the Zernike wavefront sensor has been investigated as an option for maintaining the primary mirror alignment \cite{2011ApOpt..50.2708V, 2016A&A...592A..79N, 2020OExpr..2812566C} and has been demonstrated on-sky \cite{2022ApJ...932..109V, 2024ApJ...967..171S}. 
The Pyramid Wavefront Sensor (PyWFS), as a wavefront sensor with a larger linear range at the cost of lower sensitivity \cite{2013A&A...555A..94N}, has also been investigated as a sensor to phase segmented mirrors, 
including lab tests where a PyWFS 
senses the piston, tip, and tilt modes of the primary segments \cite{2003SPIE.5169...72E}, controls them in closed loop \cite{2005OptL...30.2572E}, and implements techniques to solve wavelength phase ambiguities \cite{2006SPIE.6267E..2YP}.

However, adding any new dedicated wavefront sensor for tracking the segment phase will come with the trade-off of increasing the optical complexity of the wavefront control system, while potentially incurring throughput and sensitivity impacts to science observation. 
This work analyzes the counterpart of this trade-off by exploring the capabilities of tracking the segment phase without increasing the optical complexity of the system by using sensors already in use. 

In this two-part paper, we consider using a preexisting AO system's PyWFS to 
additionally sense the phase of the primary segments. By monitoring the phase of the primary with the existing wavefront sensor, we effectively incorporate the primary into the AO system as an auxiliary deformable mirror operating at a low temporal bandwidth. We refer to a primary mirror monitored and controlled in this way as a Controllable Segmented Primary (CSP). When considering the CSP and AO systems operating simultaneously with a single wavefront sensor, we refer to the combined system as the CSP+AO system. 
We note that this setup is not fundamentally dissimilar to a two-stage AO system operating as a woofer and tweeter, but as the temporal bandwidths, Fourier domains, and roles in the observatory as a whole differ greatly from each other, we approach this work from first principles.

This paper details the approach for accurately measuring the phase of the CSP and distinguishing it from the phase of the AO system. 
In Sec. \ref{sec:CSP_setup}, we discuss our methodology for modeling the combined CSP+AO system and disentangling the two phases. 
In \S\ref{sec:CSP_sensitivity_method}, we describe the metrics we use to measure the sensitivity and precision in the measurement of the CSP phase. 
In \S\ref{sec:CSP_results}, we present the results of these simulations, which are then discussed in \S\ref{sec:CSP_discussion}. In the companion paper (part two), we present simulations of a CSP+AO system in closed-loop operation, along with analysis of open-loop telemetry data from on-sky observations with known CSP surface aberrations.

\section{Conceptual design of a CSP+AO system} \label{sec:CSP_setup}
The most simple form a closed-loop adaptive optics system can take is a single-conjugate system with one deformable mirror and one wavefront sensor closing the loop on a natural guide star. This simple system should feature high measurement accuracy in the linear range of the wavefront sensor and should be sensitive to faint stars. 
In this work, where we introduce our methodology for incorporating the CSP into the AO system, our goal is to demonstrate that we meet these basic requirements of simple AO control while ideally maintaining the alignment and stability of the primary mirror within 
the original requirements for the primary control system. 
We choose to impose an additional requirement that the active monitoring of the primary using the wavefront sensor must remain non-invasive, meaning it must not interfere with the operation of the other control loops of the AO system. To measure the phase of the CSP without this non-invasiveness requirement, we would either need to interrupt AO operation for the measurement, or we would risk potentially introducing an unstable feedback loop between the two systems. 
Our first step to begin meeting these requirements 
is to demonstrate the ability of a PyWFS to sense and control phase modes from the CSP while ignoring any interactions with the deformable mirror.

The CSP can be controlled similarly 
to how a deformable mirror is controlled by an AO system:
The piston, tip, and tilt modes that each hexagonal mirror segment can actuate imparts unique, linearly independent phase aberrations in the pupil onto the wavefront. The segments of a primary mirror may generally have more degrees of control, but we limit the investigation in this work to the piston, tip, and tilt modes of each segment.
We use the piston-tip-tilt basis for its accessible implementation and illustrative properties. However, we highlight that the choice of basis is inessential at this stage so long as there exists a unitary matrix that can be used to convert one basis to the other without removing or adding any degrees of freedom. 

\begin{figure}[t] 
\centering
\includegraphics[width = 0.99\textwidth]{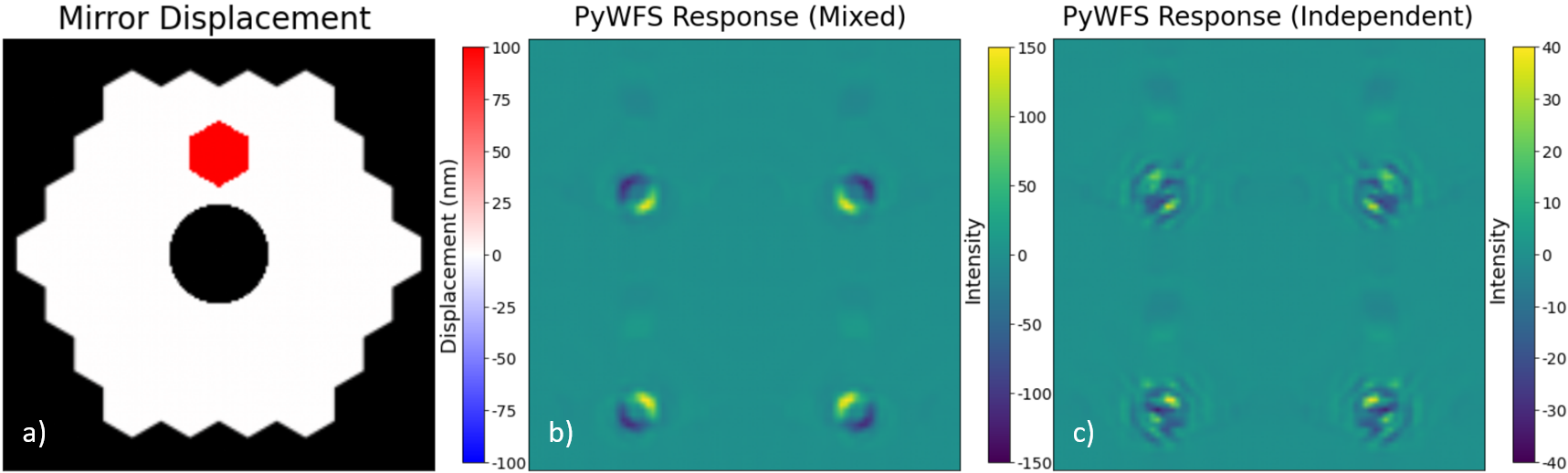}
\caption{Example of how the simulated PyWFS responds to one of the CSP modes. (a) The phase displacement of one piston mode on the CSP. (b) Measurement of this CSP mode in the mixed space of the wavefront sensor. (c) Measurement of this CSP mode in the independent space, after using Eq. \ref{eqn:CSP_ind} to remove the DM-controllable signal.
\label{fig:CSP_modes}} 
\end{figure}

Each response is then recorded by the PyWFS into an interaction matrix, $A_\text{CSP}$, where each column is the measurement from the wavefront sensor in response to each CSP mode. 
Figure \ref{fig:CSP_modes}(a) shows an example of one piston mode being applied to a segment, and then the resultant signal, referenced to a static 
wavefront, present on the PyWFS is 
displayed in \ref{fig:CSP_modes}(b). This figure is constructed from a simulation of the Keck-II AO using the IR Pyramid in NGS mode (KAO, for brevity)\cite{2020JATIS...6c9003B}, explained more in Sec. \ref{sec:sim_setup}. This demonstrates how the PyWFS is capable of sensing the modes of the CSP when isolated 
from the DM. However, the PyWFS response to these CSP modes is not independent of the PyWFS responses to the DM, creating 
the potential for signals from the DM to affect 
the CSP measurements and become invasive.

\subsection{Independent Wavefront Space} \label{sec:CSP_ind}

The CSP operates on a lower temporal bandwidth than the adaptive optics system, further motivating the non-invasiveness requirement. 
We consider the systems non-invasive when the signals from the PyWFS that generate commands for the DM are orthogonal to those for the CSP and the cross-talk between the two systems is minimized. 
We can quantify the cross-talk between the two adaptive systems by generating a matrix of overlap integrals between the PyWFS output signals from each degree of freedom of the combined system. We define the combined interaction matrix of both systems, $A_{\text{tot}}$, as a concatenation of the CSP interaction matrix, $A_\text{CSP}$, and the DM interaction matrix, $A_\text{DM}$. With each column vector continuing to be a wavefront sensor response to a mode, $A_{\text{tot}}$ then contains the response of the wavefront sensor to all of the degrees of freedom for the CSP and DM at once. 

\begin{figure}[t] 
\centering
\includegraphics[width = 0.99\textwidth]{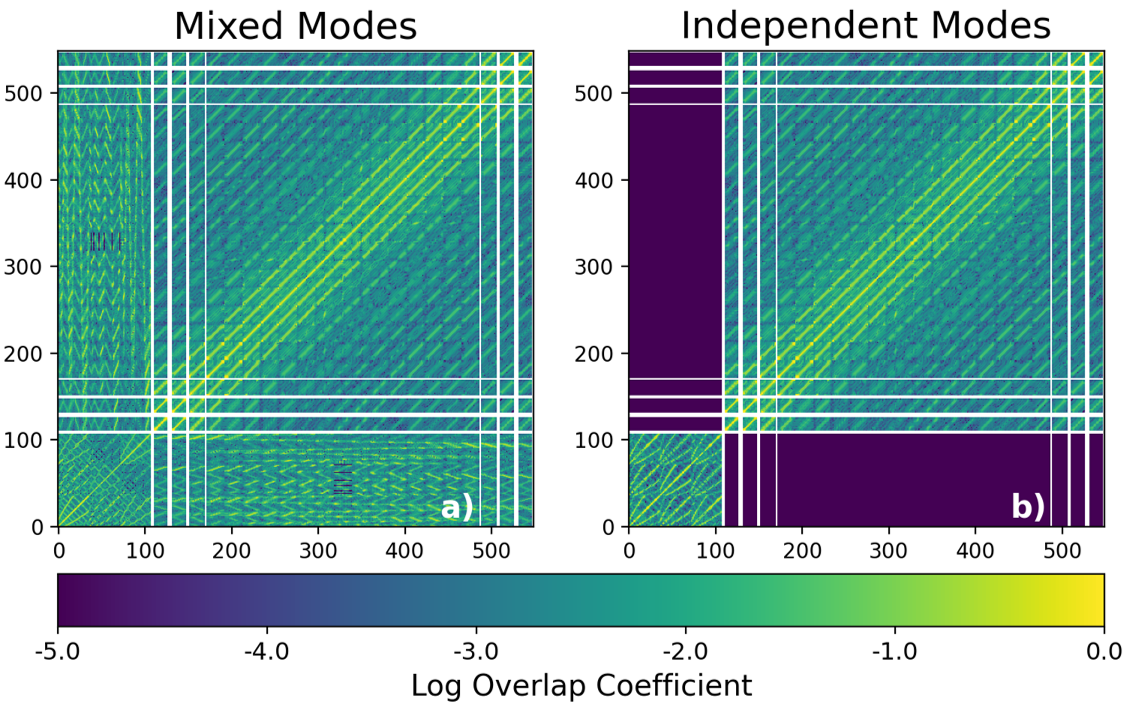}
\caption{A matrix of overlap integrals between the responses on the PyWFS for the combined DM + CSP system in log space. These integrals show the amount of similarity between the signal produced on the wavefront sensor by the deformable mirror (mode numbers 108-548) and either (a) the mixed CSP modes or (b) the independent CSP modes (mode numbers 0-107, both). } \label{fig:cross_talk} 
\end{figure}

We can quickly find the overlap integrals of the cross-talk in $A_\text{tot}$ as
\begin{equation} 
  C = A_{\text{tot}}^\intercal \times A_{\text{tot}},
  \label{eqn:CSP_crosstalk}
\end{equation}
where $A_{\text{tot}}^\intercal$ is the transpose of $A_{\text{tot}}$, meaning that $C_{jk}$ is the dot product between the WFS responses to the $j^\text{th}$ and $k^\text{th}$ degrees of freedom. We normalize the $A_\text{tot}$ matrix such that for any column vector, $\vec{A_j}$, the $j^{\text{th}}$ degree of freedom of the system, $||\vec{A_j}||^2 = 1$. We plot an example of this cross-talk of the combined CSP and DM modes from our simulation of the KAO system is Fig. \ref{fig:cross_talk}a. 
This plot can be interpreted in three sections: the CSP-CSP overlap in the bottom left corner, the DM-DM overlap in the upper right corner, and the CSP-DM overlaps in the bottom right and upper left. The vertical and horizontal bands in the plot correspond to DM actuators that fall outside of a simulated Keck aperture mask.

From Fig. \ref{fig:cross_talk}a, it is clear that the controllable space of the CSP considerably overlaps with the controllable space of the deformable mirror, as the overlap integrals in the CSP-DM overlap region are comparable in magnitude to the overlap integrals of both optics' respective individual regions. 
Violating the non-invasiveness requirement motivates us to take steps to actively remove the cross-talk between the signals that both systems generate in the wavefront sensor. 

The original signals from the PyWFS in response to a CSP degree of freedom -- which we will go on to refer to as the "mixed" signal -- is composed of two components: 
signal that can generate commands to the DM, and signals that are outside the controllable space of the DM, which we refer to as the "independent signal." 
To achieve 
non-invasiveness, we must remove the DM-controllable signal and build an interaction matrix for the CSP that is comprised of only the independent signals. We can identify the DM-controllable signal in the mixed CSP response by using the control matrix from the DM, $A^{-1}_\text{DM}$, to calculate the state of the DM that would generate the DM-controllable component to the mixed CSP signal. By multiplying the resulting DM state vector with the interaction matrix of the DM, we can project the mixed CSP signal into exclusively the DM-controllable signal. By acquiring the DM-controllable signal, we can subtract it from the mixed PyWFS response to the CSP, leaving the signal from the CSP that is independent of the DM and can only be uniquely controlled by the CSP. 
Mathematically, this method can be described as
\begin{equation}
  A_{\text{iCSP}, j} = A_{\text{CSP}, j} - \sum_{k=0}^{n_\text{acts}} c_{jk} A_{\text{DM}, k},
  \label{eqn:CSP_ind}
\end{equation}
where $A_{\text{iCSP}, j}$ is the independent component of the signal on the PyWFS from the $j^\text{th}$ degree of freedom of the CSP, 
$A_{\text{CSP}, j}$ is the mixed signal from the $j^\text{th}$ mode of the CSP, $A_{\text{DM}, k}$ is the signal from the $k^\text{th}$ actuator of the $n_\text{acts}$ DM actuators, and $c_{jk}$ is the value that projects $\text{DM}_k$ onto $\text{CSP}_j$. $c_{jk}$ is found as 
\begin{equation}
  c_{jk} = \sum_n A^{-1}_{k, \text{DM}} A_{\text{CSP}, j}.
  \label{eqn:c_jk}
\end{equation}
The sum over $n$ is a sum over the pixels of the PyWFS. An example of this independent signal is seen in panel (c) of Fig. \ref{fig:CSP_modes}.

In matrix form, this can be written as 
\begin{equation}
  A_{\text{iCSP}} = A_{\text{CSP}} - A_{\text{DM}} \times A_{\text{DM}}^{-1} \times A_{\text{CSP}}.
  \label{eqn:CSP_ind_matrix}
\end{equation}
Qualitatively, as we're multiplying the DM 
interaction matrix by its pseudo-inverse, the matrix multiplication projects the mixed wavefront sensor responses into the DM-controllable wavefront sensor space. By subtracting this projection from the CSP's mixed interaction matrix, we are left with the CSP interaction matrix comprised of only the signal that is independent and orthogonal to the DM interaction matrix. A natural consequence of this projection is that many low-order global modes that can be controlled by the primary are effectively removed in the projection into the independent space; however, these phase aberrations are still then controlled by the AO system, so their absence from the CSP loop won't affect the wavefront quality. 
After removing the DM-controllable signal from each mixed signal, we now have a CSP interaction matrix comprised only of the independent signals
, $A_\text{iCSP}$, which will be agnostic to the state of the DM when measuring the alignment of the CSP.

Despite being independently developed for distinct intended functions, 
this methodology is mathematically equivalent to the distributed modal command methodology presented by Ref. \citenum{2007ApOpt..46.4329C}, who separate the controllable spaces to control a two-DM AO system. We have redeveloped the mathematics for our specific context, where we emphasize that there isn't a clear Fourier-domain separation of the controllable spaces of the CSP and DM systems and that the two must inherently be controlled on different temporal bandwidths, but we still recognize and highlight the equivalencies in our end result.

When we examine the cross-talk between these iCSP signals 
and the DM modes in the absence of noise as in Fig. \ref{fig:cross_talk}b, we see that the CSP-DM overlap region is entirely removed. 
Inside this region, 
the strength of the cross-talk 
has been reduced down to the numerical precision limit of our calculation. 
This indicates that the 
iCSP signals are now orthogonal to the controllable space of the DM
Additionally, while we can see that the CSP-CSP overlap region has changed slightly, the DM-DM overlap region has remained unchanged. This indicates that the separation of the controllable spaces of the DM and CSP through this methodology does not interfere with the control of the DM, sufficiently meeting our non-invasiveness requirement.

\subsection{Simulation Setup} \label{sec:sim_setup}
The mathematical framework in the previous section is primarily presented in a manner separated from the specific architecture of the CSP+AO system. This is done to provide structure to future investigations for other wavefront sensors or a more general AO design. 
Moving forward, we will now describe our simulation of the combined system explicitly using a single-conjugate AO system driven in closed loop by a PyWFS, modeled after the \tcr{KAO system.} 
Using a specific AO system allows us to implement specific tests to gauge the capabilities and advantages of the CSP method. 
The full simulation of the KAO system --- built using HCIPy \cite{por2018hcipy} --- is provided in more detail in part two. The most relevant details to the investigation of the precision of the CSP are that the PyWFS samples the pupil with 44 pixels across each sub-pupil and that the DM spans 21 actuators across the pupil.

While, as implemented, the KAO system uses these sub-apertures to calculate the slopes of the wavefront in the $x$ and $y$ directions to generate commands for the deformable mirror \cite{2020JATIS...6c9003B}, we have opted to directly use the image output referenced to a \tcr{static} 
wavefront to calculate commands. Both techniques retain different advantages \cite{2022SPIE12185E..1BB}, but there is a larger domain in the form of more pixels in the PyWFS images, meaning the independent subspace is larger in the images method compared to the slopes method. This facilitates calculation of accurate iCSP commands using images, compared to using slopes.

\section{The Precision Limit} \label{sec:CSP_sensitivity_method}

In the conversion from the mixed wavefront sensor space to the independent space (Eqn. \ref{eqn:CSP_ind}), it is clear that a significant amount of the wavefront sensor response is subtracted from the measurement of the CSP amplitudes by the projection into the independent space. To address the sensitivity and brightness limit requirements for the CSP, 
we will need to generate an estimate for how precisely we can measure the CSP modes in realistic operation. 
This requires the determination of three parameters that govern the accuracy and precision of the CSP system: the responsivity, the noise sensitivity, and the susceptibility to systematic error. 


The "responsivity" is the term that we use to describe the magnitude of signal that is collected by the wavefront sensor as a result of phase error. %
\tcr{We define the signal of the PyWFS through the return-to-reference operation, introduced by Ref. \citenum{2016Optic...3.1440F},} 
\begin{equation}
    \Delta I(\phi) = I(\phi) - I_\text{ref},
\end{equation}
where $I(\phi)$ is the normalized pixel intensities of the PyWFS for an incident wavefront with phase $\phi$, and $I_\text{ref}$ is the normalized pixel intensities in response to \tcr{the static} 
wavefront. 

We consider two versions of the responsivity in this work. The "total responsivity," $R_\text{tot}$, and the "specific responsivity," $R$. 
In the linear range of the detector, the magnitude of the signal on the detector is proportional to the amplitude of phase aberration in the CSP. The total responsivity is then the coefficient to equate the two, which can be seen as: %
\begin{equation}
    \gamma_\text{sig} = R_\text{tot} \cdot \sigma_\phi, \label{eqn:sense_to_sig}
\end{equation}
where $\gamma_\text{sig}$ is the magnitude of the signal on the detector, found by summing the absolute value of the differential signal, $\gamma_\text{sig} = \sum{|\Delta I|} $, $\sigma_\phi$ is the amplitude of phase aberration that has been applied to a CSP mode, measured in nanometers, and $R_\text{tot}$ is the total responsivity, with units of photo-electrons per nanometer. In Sec. \ref{sec:sigma_sig}, we elaborate on the differences between the total and specific responsivity, as well as methods to characterize the values. 

\tcr{In a real system, there will be many sources of noise} which may be confused for a signal that would drive the CSP. The magnitude of noise present in a measurement of the CSP modes, $\gamma_\text{noise}$, is dependent on the CSP's noise sensitivity. 
$\gamma_\text{noise}$ is defined explicitly in sections \ref{sec:sigma_noise} and \ref{sec:noise_results}, where we also expand on the relationship between the CSP and its sensitivity to noise, including the specification of the different sources of noise and methods for characterizing each noise source. Broadly, $\gamma_\text{noise}$ describes the number of photo-electrons present in the detector read-out that result in random error. 

To accurately predict the sensitivity and \tcr{precision of alignment of the CSP,} 
we must compare these levels of signal and noise. Both the signal and noise will implicitly depend on parameters defined by the context of the observation (e.g. the sensing target brightness and integration time of the wavefront sensor). 
We compare the two by forming the signal-to-noise ratio from the number of photo-electrons present in the signal to \tcr{that} 
present in the noise: 
\begin{equation}
    \text{SNR} = \frac{\gamma_\text{sig, tot}}{\gamma_\text{noise}} = \frac{R_\text{tot} \cdot \sigma_\phi}{\gamma_\text{noise}}. \label{eqn:generic_SNR} 
\end{equation}


For illustrative purposes, we can define some minimum SNR that can define some level of confidence in the measurement of the alignment of the primary mirror. For a given observing context, there will be an amplitude of CSP mode that will generate the minimum SNR for measuring that mode. We can define this mode amplitude as the "Precision Limit" of the CSP mode. While proper control schemes can be applied to stably control wavefront correctors at any given SNR, the Precision Limit can provide an expectation of how the alignment of the primary will depend on target brightness, integration time, etc. By rearranging Eq. \ref{eqn:generic_SNR}, the Precision Limit takes the form 
\begin{equation}
    \sigma_\phi = \frac{\text{SNR} \cdot \gamma_\text{noise}}{R_\text{tot}}.
    \label{eqn:Precision_limit}
\end{equation}
The precision limit of a mode then acts as the expected deviation of that mode from \tcr{ideal} 
alignment. 

Implementation in a real system will also be subject to many sources of bias and other systematic errors, which will induce errors in the measurement of the CSP modes from non-zero offsets to the phase. These include non-linear effects in the WFS, residual atmospheric turbulence, aliasing effects, NCPAs, calibration errors, etc. The projection of signal into the independent wavefront sensor space does not introduce new systematic errors into the CSP+AO system and does not affect calibration errors or NCPAs, so we do not address them in this paper and assume that they are well calibrated. The sensitivity to residual atmospheric turbulence has been initially discussed in Ref. \citenum{2023SPIE12680E..0CC}. 
In this work, we bring attention to the regimes where non-linearities and aliasing may affect the accuracy of measuring the CSP phase modes, but it is beyond the scope of this paper to discuss methods for mitigating these biases. 


\subsection{Responsivity} \label{sec:sigma_sig}

We define the specific responsivity, $R$, as the quantity that directly relates the amplitude of a CSP mode to the magnitude of the signal measured by the wavefront sensor detector in response. This parameter is intrinsic to the specific wavefront sensor and architecture used for sensing phase aberrations and must be \tcr{determined in simulation.} 
To find a value that is more useful for calculating the precision limit, we must take into account 
factors such as system throughput, operating wavelength, integration time, etc. as these will also affect the magnitude of the signal measured by the wavefront sensor for a constant phase of the incoming wavefront. 

To quantify the responsivity of the CSP in the independent wavefront space, we 
build a semi-analytic model to simulate the CSP in realistic operating scenarios. 
As was described in Sec. \ref{sec:sim_setup}, we design our model to reflect the KAO system, 
using a PyWFS in the \textit{H}-band \cite{2020JATIS...6c9003B}. 
Adopting a nominal operation rate for the KAO PyWFS 
of 1\,kHz and using a $\Delta \lambda = 150$\,nm bandpass filter on an aperture with $A=6.4\times 10^5$\,$\text{cm}^2$ of collection area, we would expect a photon arrival rate of $8.6\times 10^9$\,$\gamma s^{-1}$ from a star with apparent brightness $m_H = 0$ %
across the whole primary \cite{1998A&A...333..231B}, 
which is then decreased by the throughput from the \tcr{primary mirror through the} 
PyWFS, assumed to be $\epsilon \approx 10\%$ \cite{2020JATIS...6c9003B}. Additionally, we only need to consider the arrival rate of the photons that will generate a photo-electron, so we also multiply by the quantum efficiency of the detector, which we conservatively assume to be 0.7 to match the detector used in KAO. 
Using this information, we can define a total responsivity, $R_\text{tot}$, that will allow us to estimate how many photo-electrons will be present on the detector from an individual CSP mode for a given set of wavefront sensing parameters: 
\begin{equation}
    \begin{split}
    R_\text{tot} & = R \times \left(\frac{t_{\text{int}}}{1 \text{ms}}\right) \times 10^{-m_H / 2.5} \times \frac{\text{QE}}{0.7}  \\
    & \times \frac{\epsilon}{0.1} \times \left(\frac{\Delta\lambda}{150 \text{nm}}\right)  \times \left(\frac{A_{\text{prim}}}{6.4\times 10^5 \text{cm}^2}\right),
  \label{eqn:specific_sense_signal}
  \end{split}
\end{equation}
where $R_\text{tot}$ and $R$ are the total and specific responsivities, respectively, given in units of $e^- / \text{nm}$. $m_H$ is the apparent brightness of the target that is being observed with the wavefront sensor in \textit{H}-band, and $t_\text{int}$ is the integration time of the CSP measurement. Because the CSP operates on timescales longer than the AO system, the CSP measurements are integrated over many frames of the wavefront sensor output, which 
defines the effective integration time of the CSP system --- described more in part two. As the wavelength bandpass, primary aperture collecting area, quantum efficiency, and throughput are fixed parameters, we will set those terms to their reference values for the remainder of this work. 

We can use the referenced photon arrival rate to quantify 
\tcr{the} total responsivity. As was stated in Eq. \ref{eqn:sense_to_sig}, in the linear range of the wavefront sensor, the magnitude of the signal on the detector of the wavefront sensor depends on the amplitude of the CSP mode. By varying the amplitude of our CSP modes in simulation, we can calculate $R_\text{tot}$ by measuring the corresponding magnitude of signal and finding the slope of that relationship for a given integration time and target brightness. 
This done by expanding and rearranging Eqs. \ref{eqn:sense_to_sig} and \ref{eqn:specific_sense_signal}: 
\begin{equation}
    R_\text{tot} = \frac{\gamma_\text{sig}}{\sigma_\phi \cdot t_\text{int, ms}} \times 10^{m_H / 2.5}.
    \label{eqn:responsivity_experiment}
\end{equation}
To measure the responsivity through the linear range of the PyWFS, we simulate a noiseless environment with an \textit{H}-band magnitude 0 target and actuate each individual CSP mode to amplitudes between 0.016\,nm to 530\,nm of P-V surface aberration and record the PyWFS output at each amplitude for a 1\,ms exposure time. \tcr{The piston mode raises the entire segment by the physical distance of the amplitude, while the tip and tilt modes move one edge of the segment by the given amplitude while lowering the other edge by an equal amount.} 
The total signal in each frame is the sum of the absolute value of the return-to-reference signal from each PyWFS output, which is then used as $\gamma_\text{sig}$ in Eq. \ref{eqn:responsivity_experiment}, allowing us to calculate the responsivity at every mode amplitude.


Finally, as the total responsivity will strongly depend on the context of the used AO architecture, 
it will be useful to have a metric that describes the amount of signal removed in the projection to the independent space, as this will be a relevant value in any AO system incorporating a CSP into its control scheme. 
To this end, we define the "projection efficiency", $\eta$, to describe specifically how the strength of the signal changes in the projection. We define the projection efficiency as 
\begin{equation}
    \eta = \frac{R_\text{ind}}{R_\text{mixed}},
    \label{eqn:proj_efficiency}
\end{equation}
which is the comparison of the responsivity of the CSP in the independent space, $R_\text{ind}$ to that of the mixed space of the wavefront sensor, $R_\text{mixed}$. 
We numerically evaluate the total responsivity and projection efficiency in Sec. \ref{sec:responsivity_results}.

\subsection{Noise Sensitivity} \label{sec:sigma_noise}

The noise in the system consists of many components. The Poisson statistics describing photon arrival, sky background brightness, and detector read noise and dark current 
are all sources of uncertainty in the measurement of wavefront phase. To calculate a realistic estimation of the precision limit, we will need to also estimate the magnitude of noise generated by each of these sources. The total magnitude of noise in the wavefront measurement will then equate to the sum in quadrature of each individual, independent Gaussian noise source.

As photon arrivals are governed by Poisson statistics, we can estimate the expected level of photon noise from the sensing target in simulation. Similarly to the responsivity, the sensitivity to photon noise will depend on the integration time and the brightness of the sensing target. We can estimate the sensitivity to photon noise in simulation by generating a series of simulated photon distributions for each noiseless wavefront sensor image. %
The photon noise will cause deviations from the image in comparison to the reference image and generate changes in the photo-electron distribution just as how phase aberrations create signal. After projecting into the wavefront sensor space independent of the deformable mirror, the photon noise can be found as the standard deviation in the signal from the change in the number of photo-electrons in each pixel across the set of Poisson noise masks, and the total photon noise in a given independent mode is the sum in quadrature of these noise frames, weighted by the pixel weight of the independent mode. 
Mathematically, this can be represented as: 
\begin{equation}
    \gamma_\text{PN, j} = \sqrt{\sum_{k}^{N_\text{pix}} w_{jk} \Gamma_{jk}},
    \label{eqn:photon_noise}
\end{equation}
where $\gamma_\text{PN, j}$ is the expected photon noise in the $j^\text{th}$ mode of the CSP, $\Gamma$ is the variance of photo-electrons generated from the series of simulated photon distributions in the $k^\text{th}$ pixel of the wavefront sensor. The sum is weighted by the contribution of each pixel to the measurement of the $j^\text{th}$ CSP mode, 
\begin{equation}
    w_{jk} = \frac{|A_{jk}|}{\text{max}\left( |A_j| \right)}.
    \label{eqn:weight}
\end{equation}

Read noise is a property of the detector used in the AO system. In this work, we assume that the read noise is uniform across the detector. We calculate the contribution of the read noise to the total noise of the measurement through a similar weighted sum as for the photon noise: 
\begin{equation}
    \gamma_\text{RN, j} = \sqrt{\sum_{k}^{N_\text{pix}} w_{jk} \mu^2},
    \label{eqn:read_noise}
\end{equation}
where $\gamma_\text{RN,j}$ is the expected integrated read noise for the $j^\text{th}$ mode of the CSP, $\mu$ is the reported read noise per pixel of the modeled detector, and $w_{jk}$ are the same weights \tcr{from} 
Eq. \ref{eqn:weight}. 

\tcr{
The parameters describing the distributions of many of the other noise sources are constants that do not change between observations. The random component of sky background noise, for example, will depend on the sky brightness at the location of a specific observatory. 
The contributions of other noise sources will not be derived through simulation but simply reported from accepted literature values. 
}
We present the magnitudes of the noise sources and how they combine with the responsivity to generate our precision limit estimates in Sec. \ref{sec:noise_results} and Sec. \ref{sec:CSP_precision_results}.


The sensitivity of a wavefront sensor to photon noise has been commonly used as a metric to compare between different wavefront sensors \cite{2005ApJ...629..592G, 2015OExpr..2328619P, 2016SPIE.9909E..66P, 2022SPIE12185E..2TC, 2023arXiv231010889H}. 
In Ref. \citenum{2005ApJ...629..592G}, the parameter $\beta_p$ was defined as the sensitivity to photon noise of a wavefront sensor from the equation $$\Sigma = \frac{\beta_p}{\sqrt{N_\text{ph}}},$$ where $\Sigma$ is the residual phase error in the pupil plane in radians rms following a WFS measurement of a given Fourier mode, and $N_\text{ph}$ is the number of photons incident on the wavefront sensor. Although the phase aberrations that we work with are specifically the CSP modes rather than Fourier modes, 
we will report an analog to $\beta_p$ for the CSP
in Sec. \ref{sec:CSP_sensitivity_discuss} \tcr{for consistency in literature}. 

\tcr{We} will be able to use Eq. \ref{eqn:Precision_limit} to estimate the precision of our measurements of the phase of the CSP \tcr{at any desired CSP integration time, which} 
will let us compare the expected CSP alignment performance to the designed edge-sensor alignment performance. This will allow us to determine the regimes where the CSP system will perform to a desired level of accuracy for a given level of uncertainty. 
We explore these concepts more in Sec. \ref{sec:CSP_precision_results} and \ref{sec:CSP_sensitivity_discuss}.

\section{Simulation Results} \label{sec:CSP_results}
Here, we summarize our findings describing the 
utility of using the PyWFS to simultaneously drive both the CSP and the AO system. 

\subsection{Responsivity} \label{sec:responsivity_results}
\tcr{We follow the procedure from Sec. \ref{sec:sigma_sig} to quantify the change in the PyWFS signal in response to the input phase amplitude.} 
The measurements of signal are made in units of electrons for the purpose of clarity when comparing the signal with \tcr{electrical noise sources in the detector readout.} 

\begin{figure}[t]
\centering
\includegraphics[width = 0.80\textwidth]{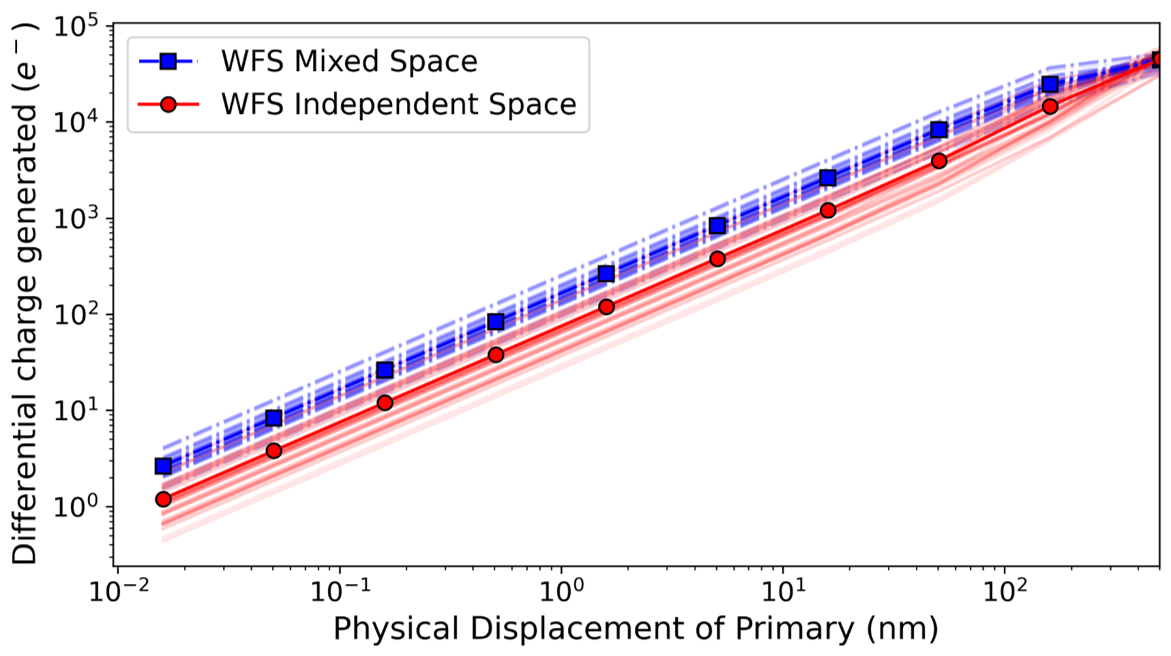}
\caption{The responsivity functions for the CSP modes on the PyWFS in both the mixed space and in the independent space. Taking the slope of these lines in the linear range gives us responsivity values of approximately 160\,$e^-$/nm in the mixed space, and approximately 75\,$e^-$/nm in the independent space. \label{fig:CSP_responsivity}} 
\end{figure}

In Fig. \ref{fig:CSP_responsivity}, we plot these measurements for each CSP mode, as well as the average across all CSP modes, showing linear behavior through $\sim 100$\,nm of surface aberration. Beyond this amplitude of surface aberration, the responsivity begins displaying non-linear behavior, as evidenced by the deviation from linearity in the figure. \tcr{This non-linear behavior is consistent with the linear range of a modulated pyramid wavefront sensor \cite{2023PASP..135k4501W, 2024A&A...681A..48C}, and will be further discussed in Sec. \ref{sec:CSP_sensitivity_discuss}.} 
We use Eq. \ref{eqn:responsivity_experiment} to calculate the responsivities as the slope from these measurements in the linear range. In blue, the mixed CSP \tcr{signals} 
show that, on average in the linear regime \tcr{of the mixed space}, the KAO PyWFS system has a total responsivity of $R_\text{mixed} \approx 160$ photo-electrons 
for every nanometer of displacement of that mode in one millisecond while observing an $m_H = 0$ star. 
In red, we see that the independent CSP \tcr{signals} 
have an average linear total responsivity of 
\begin{equation}
    R_\text{ind} = 75 \cdot \frac{\text{QE}}{0.7} \cdot \left(\frac{t_{\text{int}}}{1 \text{ms}}\right) \cdot 10^{-m_H/2.5},
  \label{eqn:specific_responsivity}
\end{equation}
with an uncertainty of $34\,e^- / \text{nm}$ due to differences between modes. \tcr{The markers are plotted to show the behavior of the mean of the segments, as opposed to the lines that show the signal generated from each individual CSP mode.} As was stated following Eq. \ref{eqn:specific_sense_signal}, the units of the responsivity are $e^- / \text{nm}$, the contributions from wavelength bandwidth, primary collecting area, and throughput have been set to their reference values, while $t_\text{int}$ indicates how many milliseconds of PyWFS images are being integrated to collect the signal, $\text{QE}$ is the quantum efficiency, and $m_H$ is the \textit{H}-band apparent magnitude of the target star. These scalings follow from the analysis in \S\ref{sec:sigma_sig} and were not tested numerically. 
These measurements of the responsivity serve as the connection between the magnitude of wavefront error caused by the CSP and the recoverable signal from the wavefront sensor.

Comparing the responsivities in the mixed and independent spaces, we arrive at a projection efficiency, defined in Eq. \ref{eqn:proj_efficiency}, of $\eta = 46\% \pm 19\%$. 
The segments in the outer ring of the Keck aperture have the highest efficiencies in each of the piston, tip, and tilt modes, at approximately $\eta \approx 60\%$, while the inner segments all have efficiencies at approximately $\eta \approx 35\%$. This is due, in large part, to limitations in our modeling of the KAO system and regularizing the DM actuators near the outer edge of the pupil. Excluding the outer segments does not significantly change our findings regarding the precision limits of the CSP. 

\subsection{Sensitivity to Noise} \label{sec:noise_results}
We now move on to characterizing the expected magnitude of noise. 
\tcr{We follow the procedure introduced in Sec. \ref{sec:sigma_noise} to quantify the photon noise component of the total noise. Equation \ref{eqn:photon_noise} allows us to convert a series of Poisson noise masks from each CSP signal to the sensitivity of that CSP signal to photon noise.}
The noise curves in the mixed space and in the independent space 
are comparative in magnitude as the photon noise spans the entire WFS space, rather than preferentially in the DM-controllable space. Additionally, the noise curves do not change with the amplitude of modal displacement as the magnitude of change in the electrons from phase aberrations is significantly smaller than the number of photo-electrons being generated in each frame. Thus, the average modal error can be expected to only depend on the integration time and target brightness. As photon noise is based on Poisson statistics, the variance scales with the integration time, quantum efficiency, and target brightness in the same way as the signal. Found through simulation, the scaling factor of the photon noise takes the form: 
\begin{equation}
    \gamma_{\text{noise, phot}} = 144\,e^- \cdot \sqrt{\frac{\text{QE}}{0.7} \cdot \frac{t_{\text{int}}}{1\,\mathrm{ms}}} \cdot 10^{-m_H / 5},
  \label{eqn:phot_noise_result} 
\end{equation}
with an uncertainty of $\pm 17\,e^-$, due to variations between modes.

Other sources of random error are the sky background, read-out noise, and dark current. 
The SAPHIRA detector, which is used in the Keck PyWFS, is known to have particularly low read noise \cite{2018JATIS...4b6001G, 2020JATIS...6c9003B}, with the Keck PyWFS measuring approximately $\sim 0.4\,e^-$ per pixel of effective read noise. We \tcr{use} 
Eq. \ref{eqn:read_noise} to find that each mode has an integrated noise of approximately $\sim 5\,e^-$. 
The photon noise approaches this magnitude of integrated noise at $t_{\text{int}} = 1\,ms$ and $m_H = 7.3$, meaning the read noise will become important for dimmer targets. 
The read noise will also increase as the square root of the number of PyWFS frames included in the effective integration time. We can incorporate the read noise into the photon noise to acquire the total noise by summing the two in quadrature.

The sky background at Mauna Kea in the \textit{H}-band has been measured to be approximately $\sim14\,\text{mag} / \text{arcsec}^2$, depending on conditions \cite{2014NewA...28...63P}, and the field of view of the PyWFS is $\sim 3\,\text{arcsec}^2$ \cite{2020JATIS...6c9003B}. This means the sky background will not be relevant unless we are using the CSP methodology on targets of \textit{H}-band brightness 12 or dimmer. 
For the purposes of this work, this implies that the read noise will always dominate over the sky background and that the sky background can be ignored while calculating the precision limit. 
The noise from the dark current of the SAPHIRA detector in operation at Keck is not reported \cite{2020JATIS...6c9003B}, and Ref. \citenum{2017AJ....154..265A} reports the noise from SAPHIRA dark currents to be significantly lower than what is seen in read noise, so we also assume that the noise associated with the PyWFS dark current is negligible in comparison to the read noise for integration times typical of AO operation.

The atmospheric turbulence profile also causes errors in accuracy when computing commands for the CSP. However, we don't treat \tcr{this} 
as a noise source for the purposes of estimating the precision of the CSP. The errors caused by the atmosphere are from physical differences in optical path length from the top of the atmosphere to the telescope. This effect does not generate uncertainty in signal that can be resolved by collecting photons from a brighter source, but it is instead a source of error in location for where photo-electrons are generated on the detector. Additionally, unlike electronic noise, increasing the number of frames of integration does not increase the magnitude of this uncertainty, but instead decreases it by allowing uncorrelated regions of atmosphere to cross the pupil. As such, while there will be cases where the accuracy of the CSP is limited by atmospheric turbulence, we do not include it in this derivation of the CSP precision limit. As mentioned before, a quantitative investigation into how the presence of the atmosphere impacts the CSP measurements is done in Ref. \citenum{2023SPIE12680E..0CC}.


Thus, we can now find the total integrated noise by summing the photon noise and read noise in quadrature. For clarity, we represent the integrated photon as 
$\gamma_\text{PN} = 144 \sqrt{\text{QE} / 0.7}\,e^-$ 
and the integrated read noise as $\gamma_\text{RN} = 5\,e^-$. The total integrated noise, with units of $e^-$, can then be reported as 
\begin{equation}
    \gamma_{\text{noise}} = \sqrt{\frac{t_{\text{int}}}{1\,\mathrm{ms}} \left(\gamma_\text{PN}^2 \cdot 10^{-m_H / 2.5} + \gamma_\text{RN}^2\right)}.
  \label{eqn:tot_noise}
\end{equation}

\subsection{Precision Limit Results} \label{sec:CSP_precision_results}

With both the responsivity and noise sensitivity estimated, we can now derive a numerical expression of the CSP precision limit, as described in Eq. \ref{eqn:Precision_limit}: 
\begin{equation}
  \sigma_\phi = \frac{\text{SNR}\cdot \sqrt{\gamma_\text{PN}^2 \cdot 10^{-m_H / 2.5} + \gamma_\text{RN}^2}}{R_\text{ind} \cdot \sqrt{t_{\text{int}}}} \cdot 10^{m_H / 2.5}.
  \label{eqn:sensitivity}
\end{equation}

\begin{figure}[t]
\centering
\includegraphics[width = 0.80\textwidth]{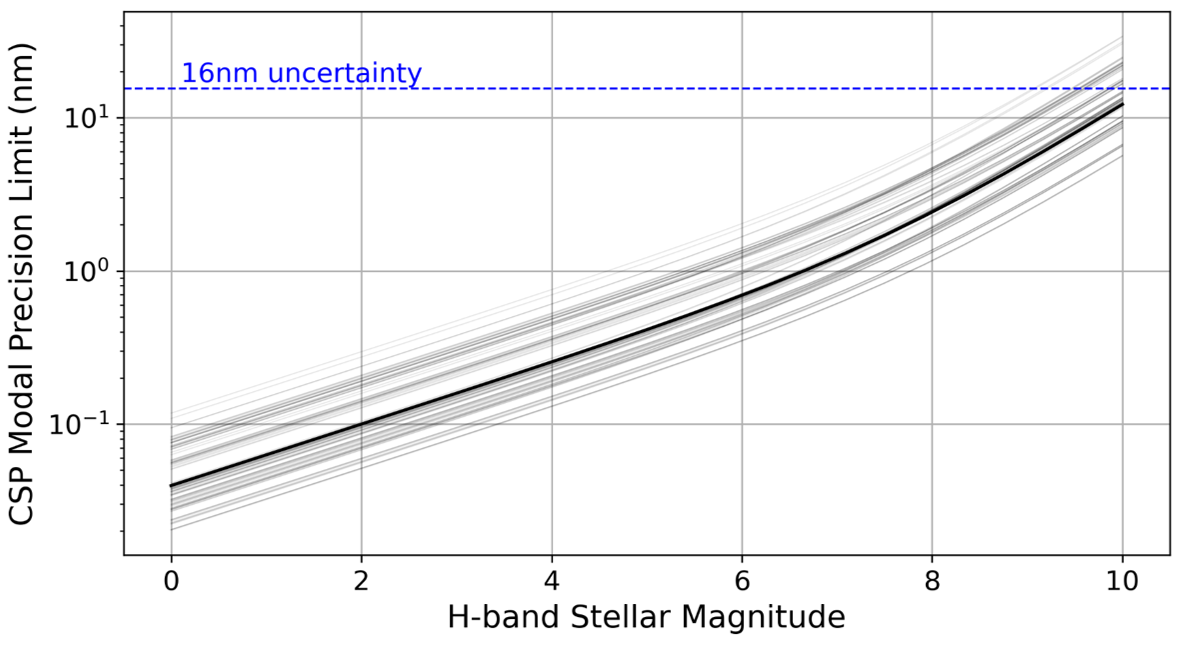}
\caption{Example of a precision limit plot showing the minimum amplitude aberration for every iCSP \tcr{signal} 
to reach the target SNR. This curve is found from Eq. \ref{eqn:sensitivity} with a target $\text{SNR} = 3$, noise described as Eq. \ref{eqn:tot_noise}, responsivity described in \ref{eqn:specific_responsivity}, and an integration time of $t_{\text{int}}=30$\,sec. With these example values, we can achieve an RMS \tcr{surface} flatness of \tcr{$<16$\,nm} 
for \tcr{every iCSP signal} 
on a star with apparent \textit{H}-band magnitude of \tcr{$m_H \leq 9$.} 
\label{fig:CSP_sensitivity}} 
\end{figure}

We can now assume typical values for each of these parameters to plot our sensitivity for each mode. In Fig. \ref{fig:CSP_sensitivity}, we 
\tcr{plot Eq. \ref{eqn:sensitivity}, assuming an example target} 
signal to noise ratio 
\tcr{as} $\text{SNR} = 3$, \tcr{the} 
photon noise, read noise, and responsivity are as calculated, and \tcr{an} 
effective CSP integration time is $t_{\text{int}} = 30\,\text{s}$ as an estimated evolution time to average out the atmospheric residuals. 
\tcr{We also plot a horizontal reference line at 16\,nm of precision. As the electrical noise from the edge sensors in the Keck primary are expected to generate an rms surface misalignment of 27\,nm \cite{1994SPIE.2199..105C}, 
by assuming that the misalignment is distributed equally between the piston, tip, and tilt modes, we can use 16\,nm as the reference for the expected sensitivity performance limit for the primary for each CSP mode.} 
We can see that for targets brighter than \tcr{$m_H \leq 9$}, 
we can sense \tcr{every iCSP signal} 
to a precision better than the edge sensor \tcr{reference}. 

Equation \ref{eqn:sensitivity} can also be used to estimate the amount of CSP integration time needed to achieve the target SNR. 
By rearranging Eq. \ref{eqn:sensitivity} to solve for the integration time, $$t_\text{int} = \left(\frac{\text{SNR}\cdot \sqrt{\gamma_\text{PN}^2 \cdot 10^{-m_H / 2.5} + \gamma_\text{RN}^2}}{R_\text{ind} \cdot \sigma_\phi} \cdot 10^{m_H / 2.5}\right)^2, $$
we can calculate the necessary integration time needed to achieve a defined precision limit, $\sigma_\phi$, as a function of sensing target brightness. This function takes a similar shape to Fig. \ref{fig:CSP_sensitivity} in log-linear space. To sense to a precision of
\tcr{$\sigma_\phi = 16$\,nm of misalignment for the median CSP mode in the independent space, }
the required integration time ranges from \tcr{$\sim 0.3$\,ms} 
for an \textit{H}-band magnitude 0 target to approximately \tcr{$\sim 40$\,seconds} 
for a target of $m_H = 10$. \tcr{These integration times are the expected amount of duration necessary to achieve the target SNR, but we do not account for biasing errors from atmospheric turbulence in this estimation.}



\section{Discussion} \label{sec:CSP_discussion}
The over-arching goal of our project is to actively support the phasing of a segmented primary mirror by \tcr{monitoring semi-static variations in the primary mirror alignment that are not observed by} 
the edge sensing systems of segmented telescopes. In this paper, we have shown an approach for using the wavefront sensor of an AO system to monitor the alignment of the primary by disentangling the phase of the DM from the independent phase. We have also, in an example modeled from the KAO system, shown that we have retained the ability to make make measurements of the phase of the CSP from the independent signal in the presence of noise. We go on to discuss additional implications from these results. 

Prior to this, it is necessary to briefly acknowledge 
that primary mirror phasing is not necessarily limited to capacitance-based edge sensors, which were initially referenced in Sec. \ref{sec: intro}. The Giant Magellan Telescope (GMT) will be composed of seven large circular primary mirror segments, whose phasing will be maintained via a \tcr{combination of an on-sky phase retrieval system and} holographic dispersed fringe sensors \cite{2016ApOpt..55..539V, 2022JATIS...8b1513H, 2022SPIE12185E..1CB}. In Ref. \citenum{2023aoel.confE..85C}, 
we investigated the application of the CSP methodology to AO systems with different actuator densities and other architecture changes. 
In this, we found that higher numbers of DM actuators across the primary mirror segments will decrease the size of the independent CSP space and decrease the accuracy of the CSP commands.
As GMT represents an extreme case of DM actuator density \tcr{per segment}, we did not consider it in any of our analysis.

\subsection{Sensitivity} \label{sec:CSP_sensitivity_discuss}

The edge sensor systems on the Keck telescopes is meant to maintain the alignment of the primary to $\sim 5$\, nm at a cadence of $\sim 2$\,Hz \cite{1994SPIE.2199..105C}, and, as seen from Fig. \ref{fig:CSP_sensitivity}, the PyWFS from the AO system reaches comparable levels of precision at cadences around $\sim 0.1$\,Hz, corresponding to effective integration times of $\sim 10$\,s, which would also depend on the sensing target brightness. At this operation rate, the PyWFS would not be capable of maintaining the alignment of the primary as a standalone sensor. We can, instead, use this methodology in tandem with the primary mirror control system to verify and update the reference positions of each mirror segment, resulting in a stable surface alignment. 
In this subsection, we \tcr{discuss} the values of responsivity, sensitivity to noise, and the precision limit as found in 
simulation. 

\tcr{In Sec. \ref{sec:CSP_precision_results}, we argued that the expected misalignment of the piston, tip, and tilt of each mirror segment would be at 16\,nm rms, assuming the residual misalignments from the edge sensor system is distributed uniformly between the degrees of freedom of each segment.} 
Our results from Sec. \ref{sec:CSP_precision_results} indicate that we can achieve this level of sensitivity for the average \tcr{CSP mode in the independent space} 
through targets of \textit{H}-band brightness of $m_H \leq 10$ for integration times on the scale of 30 seconds. Reference \citenum{2022SPIE12182E..09R} estimates that the drifting of edge sensor reference positions on the Keck-II telescope appears on the timescale of weeks to months, so through the 30-second integration time, we can assume that the alignment of the primary segments remains roughly constant. 
Further studies of mirror alignment and the active control systems correlate the segment misalignments --- in the form of terrace and focus modes --- to elevation angle \cite{2022SPIE12182E..09R, 2023ApOpt..62.5982C}. While operating the AO system in closed loop, the elevation angle of the telescope also does not appreciably change in the integration time of 30 seconds. 
As a result, we can monitor the primary aberrations correlated with the elevation 
through this method, as the reference \tcr{phase}, 
saved as a calibrated wavefront sensor readout, will not change in time.

\tcr{We investigated the precision limit to an \textit{H}-band brightness of $m_H \leq 10$.} 
This brightness is near the limiting brightness of the KAO PyWFS \cite{2020JATIS...6c9003B}, indicating that we can effectively monitor the phase of the primary for the same brightness range of stars as the wavefront sensor typically operates, achieving our brightness operation requirement. \tcr{This brightness limit is also compatible with the science cases introduced in section \ref{sec: intro}, allowing this methodology to be used in the same science cases where the PyWFS would be utilized.} 

From Fig. \ref{fig:CSP_responsivity}, we were able to observe that at phase aberrations of approximately $\sim 100$\,nm of mirror displacement, non-linear effects begin to appear in the mixed-space responsivity of the PyWFS, while the independent-space responsivity remained linear. 
This implies that phase aberrations in the DM-controllable space at this amplitude of phase aberration will also generate non-linear effects. As a result, the linear separation that we apply to disentangle the independent CSP space from the mixed space will no longer fully remove the DM-controllable signal. This will generate cross-talk between the CSP and AO systems and could cause inaccuracies in the estimation of the CSP phase, potentially affecting the closed-loop stability of the CSP control. This indicates that beyond $\sim 100$\,nm of surface aberrations is the regime where non-linear effects may begin to generate biasing errors. It may be possible to mitigate this issue by employing non-linear phase reconstruction techniques to estimate the phase incident on the wavefront sensor and then separating the resultant phase into DM-controllable phase and independent phase, but this is beyond the scope of this paper. \tcr{Surface errors on the primary mirror segments induce phase aberrations well within this non-linear regime\cite{2024SPIE13094E..3FG}, which would introduce errors into the measurement of the CSP modes. We provide further discussion of the impacts of the segment surface errors in the discussion of paper 2 with other details regarding implementing this methodology on physical systems.}

\tcr{As was stated in Sec. \ref{sec:noise_results}, we do not incorporate error from atmospheric residuals into the formulation of the precision limit.} 
However, \tcr{as the precision limit in most science cases with the PyWFS is on the scale of $\sigma_\phi \leq 15$\,nm}, there are clearly areas of parameter space where the uncertainty from the atmosphere is the limiting factor in the accuracy of the CSP commands. 
We leave further analysis of how the accuracy of CSP commands scales with integration time for following the collection of additional on-sky data.

\tcr{\subsection{Sensitivity to Photon Noise, $\beta_p$}} \label{sec:beta_p_discuss}
In Sec. \ref{sec:sigma_noise}, we referenced the $\beta_p$ parameter from Ref. \citenum{2005ApJ...629..592G}, which relates the residual phase error in the pupil plane following the measurement of a given Fourier mode and the number of photons used in that measurement, as a common metric for describing wavefront sensor sensitivity. We can use our simulation of the CSP+AO system to create a similar, effective $\beta_p$ value by comparing the expected residual phase error from our derived precision limit for a given sensing target brightness for each degree of freedom of the CSP. This was calculated by setting each individual CSP mode to the precision limit of a given integration time and target brightness in the photon noise-limited regime and then multiplying the resultant rms pupil phase by the square root of the number of photons for those conditions. 

\begin{figure}[t]
\centering
\includegraphics[width = 0.70\textwidth]{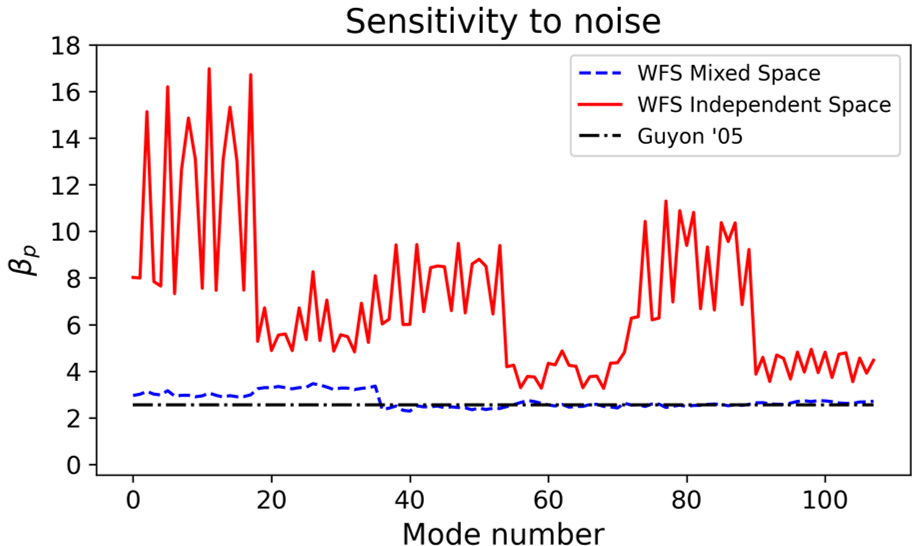}
\caption{The effect of the controllable space filtering on the sensitivity to noise. The $\beta_p$ values for the CSP in both the mixed space and the independent space show that the CSP is $2.6 \pm 1.1$ times more sensitive to photon noise than the mixed space, while the dotted black line indicates the average value of $\beta_p$ at high spatial frequencies for a modulated PyWFS. \label{fig:CSP_betaP}}
\end{figure}

In Fig. \ref{fig:CSP_betaP}, we plot the effective $\beta_p$ values for the CSP in both the mixed space and in the independent space, as well as co-plotting the $\beta_p$ derived in Ref. \citenum{2005ApJ...629..592G}, for the average value of $\beta_p$ for high spatial frequencies of the modulated pyramid wavefront sensor. In this figure, there is agreement between the effective $\beta_p$ values measuring the CSP in the mixed space and the values calculated by Ref. \citenum{2005ApJ...629..592G}, validating that the approach we used to estimate the effective $\beta_p$. We also then observe that the effective $\beta_p$ from measurements made in the independent wavefront space are approximately $2.6 \pm 1.1$ times larger than the mixed space, where we would be incapable of monitoring the primary non-invasively. This, again, indicates that measuring the phase of the primary in the independent wavefront sensor space is more susceptible to noise than in the mixed space where these measurements are not possible. 


Overall, we've shown that in many regimes where the PyWFS drives the AO system, incorporating the primary mirror to form the CSP+AO system is a viable option to help maintain the alignment of primary to within the desired requirements while also improving the performance of the AO system. 


\section{Conclusion} \label{sec:conclusion}
Up to this point, we have introduced a methodology to identify and separate the controllable spaces of two optics in the measurements of a wavefront sensor. This allows us to measure the alignment of a segmented primary mirror in simultaneous operation with an AO system without impacting the operation of the AO system. We have also shown from a model of a realistic system that, despite removing significant amounts of the signal in the projection to the independent wavefront sensor space, these independent measurements can still outperform the necessary precision requirements for relevant wavefront sensing targets. 
In part two, we \tcr{will} further develop simulations of the operation of a CSP+AO system to demonstrate that the independent measurements can control the CSP alignment in a stable manner. We \tcr{will} also show analysis of on-sky telemetry data from the KAO PyWFS that validate our simulated results. \tcr{Lastly, discussion of some details regarding practical implementation of this methodology in real systems is also provided in paper 2.}

\subsection*{Disclosures}
The authors have no relevant financial interests or other potential conflicts of interest to disclose. 

\subsection* {Code, Data, and Materials Availability} 
Data sharing is not applicable to this article, as no new data were created or analyzed. Code to recreate simulations can be made available upon request to the corresponding author. 


\subsection* {Acknowledgments}
The W. M. Keck Observatory is operated as a scientific partnership among the California Institute of Technology, the University of California, and the National Aeronautics and Space Administration. The Observatory was made possible by the generous financial support of the W. M. Keck Foundation. The authors wish to recognize and acknowledge the very significant cultural role and reverence that the summit of Maunakea has always had within the indigenous Hawaiian community. We are most fortunate to have the opportunity to conduct observations from this mountain.
This work was supported by the Heising-Simons Foundation through grant \#2020-1821.


\bibliography{CSP_pt1}   
\bibliographystyle{spiejour}   





\end{document}